\documentclass{pasj00}

\draft

\begin{document}
\SetRunningHead{Author(s) in page-head}{Running Head}

\title{Cryogenic Volume-Phase Holograpic Grisms for MOIRCS}

\author{
Noboru \textsc{Ebizuka,} \altaffilmark{1, 2}
 Kotaro \textsc{Ichiyama,} \altaffilmark{3}
 Toru \textsc{Yamada,} \altaffilmark{3}
 Chihiro \textsc{Tokoku,} \altaffilmark{3}
 Masato \textsc{Onodera,} \altaffilmark{4, 5}
 Mai \textsc{Hanesaka,} \altaffilmark{6}
 Kashiko \textsc{Kodate,} \altaffilmark{6}
 Yuka Katsuno \textsc{Uchimoto,} \altaffilmark{3, 7}
 Miyoko \textsc{Maruyama,} \altaffilmark{8}
 Kazuhiro \textsc{Shimasaku,} \altaffilmark{9}
 Ichi \textsc{Tanaka,} \altaffilmark{10}
 Tomohiro \textsc{Yoshikawa,} \altaffilmark{11, 10, 3}
 Nobunari \textsc{Kashikawa,} \altaffilmark{12}
 Masanori \textsc{Iye,} \altaffilmark{12}
 and Takashi \textsc{Ichikawa} \altaffilmark{3}
 }

\altaffiltext{1} {Plasma Nanotechnology Research Center, Nagoya University, Furo-cho, Chikusa-ku, Nagoya 464-8603} 
\altaffiltext{2} {RIKEN (The Institute of Physical and Chemical Research), Hirosawa, Wako 351-0198}\email{ebizuka@riken.jp}
\altaffiltext{3} {Astronomical Institute, Tohoku University, Aramaki, Aoba-ku, Sendai 980-8578}
\altaffiltext{4} {Institute for Astronomy, ETH Zurich, Wolfgang-Pauli-strasse 27, 8093 Zurich, Switzerland}
\altaffiltext{5} {CEA-Saclay, DSM/DAPNIA/Service dfAstrophysique, 91191 Gif-sur-Yvette Cedex, France}
\altaffiltext{6} {Department of Mathematical and Physical Science, Japan Women's University, Mejirodai, Bunkyo-ku, Tokyo 112-8681}
\altaffiltext{7} {Institute of Astronomy, The University of Tokyo, Osawa, Mitaka, Tokyo 181-0015}
\altaffiltext{8} {College of Science and Technology, Nihon University, Kanda-Surugadai, Chiyoda-ku, Tokyo 101-8308}
\altaffiltext{9} {Department of Astronomy, The University of Tokyo, Hongo, Bunkyo-ku, Tokyo 113-0033}
\altaffiltext{10} {The Subaru Telescope, National Astronomical Observatory of Japan, 650 North A'ohoku Place, Hilo, HI 69720,  USA}
\altaffiltext{11} {Koyama Astronomical Observatory, Kyoto Sangyo University, Motoyama, Kamigamo, Kita-ku, Kyoto, 603-8555}
\altaffiltext{12} {National Astronomical Observatory of Japan, Osawa, Mitaka, Tokyo 181-8588}


\maketitle

\begin{abstract}
We have developed high dispersion VPH (volume phase holographic) grisms with zinc selenide (ZnSe) prisms for the cryogenic optical system of MOIRCS (Multi-Object near InfraRed Camera and Spectrograph) for {\it Y-, J-, H-} and {\it K-}band observations.   We fabricated the VPH gratings using a hologram resin.  After several heat cycles at between room temperature and 120 K, the VPH gratings were assembled to grisms by gluing with two ZnSe prisms.  Several heat cycles were also carried out for the grisms before being installed into MOIRCS. We measured the efficiencies of the VPH grisms in a laboratory, and found them to be 70\% - 82\%.  The performances obtained by observations of  MOIRCS with the 8.2 m Subaru Telescope have been found to be very consistent with the results in the laboratory test.  This is the first astronomical application of cryogenic VPH grisms.
\end{abstract}

\section{Introduction}
Recent innovative technologies in the near-infrared have boosted the development of new instruments for astronomy.  MOIRCS (Multi-Object near InfraRed Camera and Spectrograph) for the 8.2m Subaru Telescope is one of the most powerful infrared instruments, which fully makes use of the advent of such technologies.  The multislit spectrograph, which enables us to observe more than 50 objects at a time with cooled slit masks, is a unique function among instruments for 8-10 m telescopes (Ichikawa et al. 2006; Suzuki et al. 2008).  The advantages of MOIRCS having a wide field of view and capability for multiobject spectroscopy are enjoyed by many astronomical observations for solar system, star-forming regions, nearby galaxies, and high-z galaxies in both spectroscopic and imaging modes.
   
Spectroscopic observations of MOIRCS with resolving powers (R = $\lambda/\Delta\lambda$) of R=500 and R=1300 are very efficient for faint or distant objects to obtain the redshifts of galaxies (Hayashi et al. 2009; Yoshikawa et al. 2010), for example.  However, atmospheric emissions of the OH radical, which are main noise sources in near-infrared, e.g., {\it Y-, J-} and {\it H-}bands, often hide or disturb the emission or absorption features of targets.  To avoid OH emission forest, a higher resolving power with R$\geq$2500 is strongly demanded, so that more spectroscopic features would fall between the OH lines with a higher probability.  A higher resolving power is also desirable for the analysis of line profiles of the distant galaxies, for instance, to study the physical condition, internal kinematics, line ratio, and so forth.  If lines of a low dispersion spectrum, such as the redshifted [NII] (rest-frame wavelength: 654.8nm) and H$\alpha$ (656.3nm), are overlapped with each other, deblending of the overlapped spectral lines of faint distant galaxies is a quite difficult task due to their low signal to noise ratio expected within some handy exposures.  Moreover, the higher resolving power will help us to understand the physical and environmental conditions of a quite distant gamma-ray burst through analysis of the absorption-line profiles in the optical afterglow (Kawai et al. 2005; Totani et al. 2006).
   
Direct vision optics with all lenses of MOIRCS does not give a space large enough for a high-dispersion element, like an echelle grating.  Since a VPH (Volume-Phase Holographic) grism has many advantages over conventional surface-relief gratings, it is reasonable choice as a higher disperser.  The VPH grating has a layer with sinusoidal modulation of the refractive index.  The VPH grism consists of a VPH grating sandwiched between two prisms so that a specific order and wavelength can be aligned in the direct vision direction.  The VPH grating achieves a very high diffraction efficiency up to 100 \% for s- or p-polarized light at a specific wavelength, when the grating of refractive-index modulation and the electromagnetic wave are strongly coupled by adjusting the amplitude of the refractive-index modulation, the thickness of the layer, and incident angle of the light wave (Kogelnik 1969; Diskson et al. 1994; Barden et al. 1998, 2000; Rallison et al. 2003; Oka et al. 2003).

Many VPH gratings are currently used in various spectrographs for visible wavelengths (Ebizuka et al. 2011).  They will also be useful for near-infrared spectrographs if the performance at low temperature is satisfactory (Tamura et al. 2003).  In this paper, we describe the development of VPH grisms with high efficiency for {\it Y} ($\lambda_{center}=1.02$\micron),  {\it J} (1.25\micron),  {\it H} (1.65\micron), and {\it K} (2.2\micron) bands, which are now in use for MOIRCS at cryogenic temperature ($\sim$100 K) (Ichikawa et al. 2008; Nakajima et al. 2008).  Examples for astronomical observations are also presented.

\begin{figure}
  \begin{center}
    \FigureFile(59.91mm, 40mm){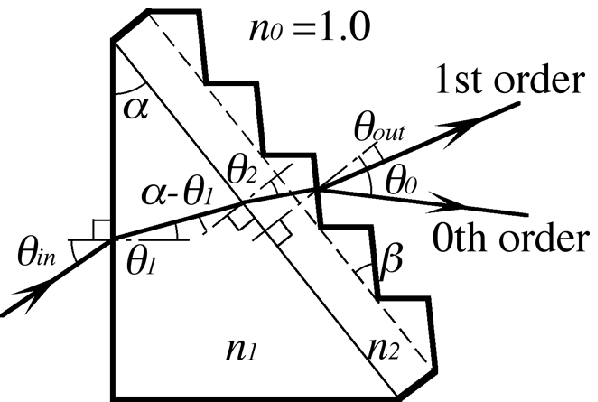}
    \FigureFile(64.58mm, 40mm){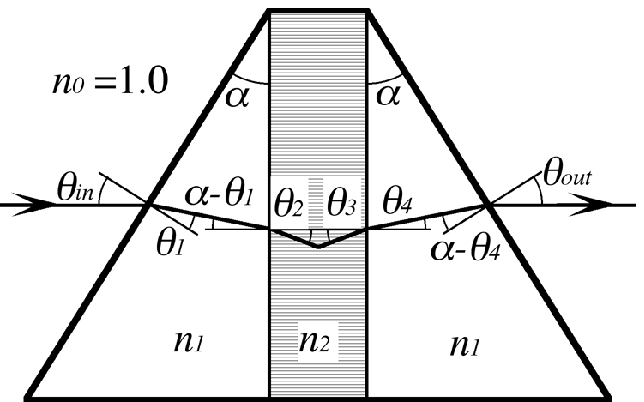}
   \end{center}
  \caption{Schematic representation of a replicated surface relief grism (left) and VPH grism (right).}\label{fig.1}
\end{figure}

\section{Grism with High Index Prism}
A prism with larger apex angle, $\alpha$, and higher refractive index, $n$, provides higher dispersion for a grism because the dispersion of a grism is proportional to optical path difference between the top and bottom rays in the prism (Ebizuka et al. 1998).  If a surface-relief grism consists of a high-index prism such as is made of zinc selenide (ZnSe, $n=2.45$ at 1.5\micron) and a replicated grating as shown in the left panel of figure \ref{fig.1}, the dispersion of the grism is limited by the critical angle at the boundary between the prism and the replica.  When the prism and replicated grating are $n=2.45$ and $n=1.5$, respectively, for the surface-relief grism, and a ray enters vertically in the prism ($\theta_{in}= 0,\ \alpha - \theta_1= \alpha$), the critical angle is $37.8^{\circ}$.  In the case of the surface-relief grism, the prism should be $\alpha>35^{\circ}$.  If a ray enters obliquely in the surface-relief grism from the prism side, as shown in the left panel of figure \ref{fig.1}, the critical angle becomes larger, but the optical axis of the grism is sifted in parallel between the incident and diffracted beams.  The shift of optical axis becomes a cause of vignetting in a following optical element.

On the other hand, the limitation of critical angle for a VPH grism, as shown in the right panel of figure \ref{fig.1}, is less strict than that of the surface-relief grism.  When prisms and a VPH grating are $n=2.45$ and $n=1.5$, respectively, for the VPH grism, the critical angle for $\alpha-\theta_1$ is $37.8^{\circ}$.  In the case of the VPH grism, a prism should be $\alpha<58^{\circ}$.  Furthermore, the VPH grism has two prisms, and a VPH grating achieves high dispersion with very high diffraction efficiency.  This indicates that a VPH grism has capability for higher dispersion applications compared with a surface-relief grism (Ebizuka et al. 2003).
 
\begin{figure}
  \begin{center}
    \FigureFile(110.78mm,75mm){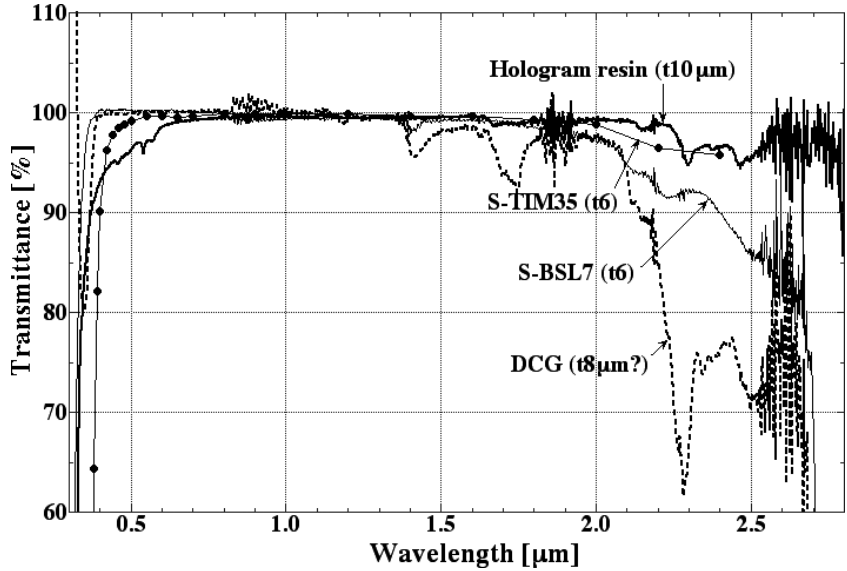}
   \end{center}
  \caption{Transmission of a hologram resin, dichromated gelatin (DCG) and optical glasses (S-BSL7 and S-TIM35).}\label{fig.2}
\end{figure}

\section{Fabrication of VPH Grisms}
We use a hologram resin as a recording material for VPH gratings supplied by Nippon Paint Co. Ltd.  The hologram resin is transparent up to 2.7\micron, except for small absorption at between 2.1 and 2.5\micron, although dichromated gelatin (DCG), which is another material for a volume hologram, has small absorption at 1.4 and 1.7\micron, and moderate absorption above 2.1\micron\  (figure \ref{fig.2}).  The resin consists of a radical polymerized monomer (RPM), a cation polymerized monomer (CPM), a dye, a solution, and a few other trace elements (Kawabata et al. 1994).

Two kinds of optical glass synthesized by Ohara Inc. are used for substrates of hologram plates for MOIRCS VPH grisms: S-BSL7 for the {\it Y-, J-} and {\it H-}bands, and S-TIM35 for the {\it K-}band.  These glass substrates have comparable coefficient of thermal expansion to the ZnSe prism.  S-TIM35 is transparent up to 2.4\micron, while S-BSL7 exhibit weak absorption above 1.8\micron\  (figure\ref{fig.2}).  The size of the glass substrate is 70 mm by 70 mm, and 3 mm in thickness.  The resin with glass beads is spread on a glass substrate, and is covered with another glass substrate after heating for vaporization of the solution in the resin.  Therefore, the thickness of the resin's layer is controlled by the diameter of glass beads as a spacer.  

\begin{figure}
  \begin{center}
    \FigureFile(98.9mm,30mm){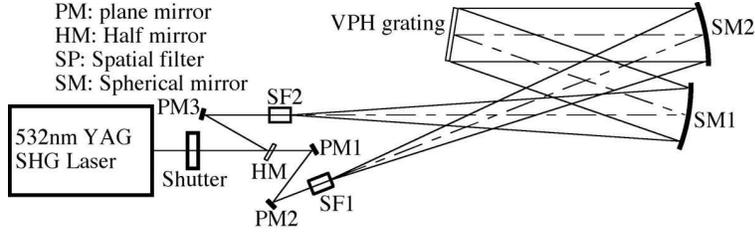}
   \end{center}
  \caption{Schematic representation of an exposure optics of a two-beam interferometer with spherical mirror collimators for a holographic grating .}\label{fig.3}
\end{figure}

\begin{table}
  \caption{Designed Specifications of MOIRCS VPH grisms.}\label{tab.1}
  \begin{center}
    \begin{tabular}{lclclclc|c|c|c|}
      \hline
      & Nominal central & Wavelength & Resolving & Modulation & Grating & Vertex angle \\
      Band & wavelength & coverage  & power\footnotemark[$*$] & frequency & thickness & of prism \\
      & [\micron] & [\micron] &  ($\lambda/\Delta\lambda$) & [periods/mm] & [\micron] & [$^{\circ}$] \\ \hline
      { \it Y} & 1.025 & 0.95$\sim$1.10 & 3150 & 1016 & 20.0 &20.0\\
      { \it J} & 1.25 & 1.15$\sim$1.34 & 3000 & 813 & 29.5& 19.8\\
      { \it H} & 1.65 & 1.52$\sim$1.78 & 2950 & 618 & 40.4 & 20.0\\
      { \it K} & 2.20 & 2.00$\sim$2.40 & 2640 & 424 & 30.0 & 18.5\\
      \hline
       \multicolumn{4}{@{}l@{}}{\hbox to 0pt{\parbox{85mm}{\footnotesize    
     \par\noindent
     \footnotemark[$*$] Calculated for 0".5 slit.
    }\hss}}
    \end{tabular}
  \end{center}
\end{table}

Figure \ref{fig.3} shows a schematic representation of a two-beam interferometer with a 532 nm green laser.  The optical layout cancels aberrations of the two spherical mirrors as collimators.  The RPM polymerized by UV and visible light has a higher refractive index, while the CPM polymerized by UV light has a lower refractive index.  When bright and dark stripes of interferometric laser modulation are formed onto a hologram resin plate by the interferometer, RPM is polymerized.  As the concentration of RPM decreases in bright parts, RPMs move from dark to bright parts and CPMs move from bright to dark parts.  Then, the refractive-index modulation of a volume phase hologram is formed by interferometric laser exposure, namely, the development of the hologram processes simultaneously under laser exposure.  While both resins are polymerized, the refractive index modulation is fixed by UV light exposure after laser exposure.  It is noted that the development of DCG is a wet process, while that of the hologram resin is a dry process.  The VPH grating and two ZnSe prisms are assembled into a VPH grism by an optical adhesive.  Since MOIRCS has two array detecters, a pair of grisms are necessary for each band.  The specifications of MOIRCS VPH grisms are given in table \ref{tab.1}.

\begin{figure}
  \begin{center}
    \FigureFile(109.8mm,75mm){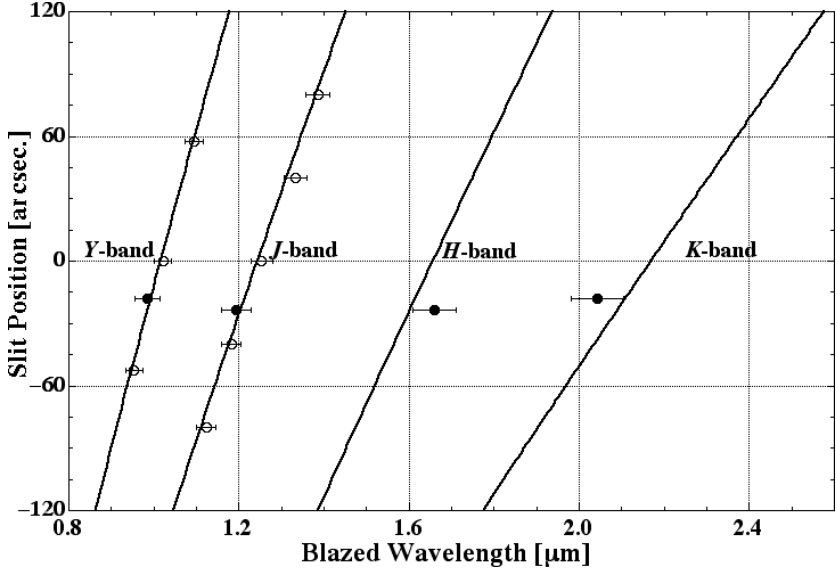}
   \end{center}
  \caption{Slit position versus blazed wavelength.  The solid lines indicate calculations in which the incident and diffracted beams in the VPH gratings satisfy the Bragg condition.  Open circles represent the wavelength of the peak efficiencies taken by the spectrophotometer, and filled circles represent the wavelength of the peak efficiencies obtained by observations of standard stars with the Subaru Telescope and MOIRCS.}\label{fig.4}
\end{figure}

\section{Performance Verification for VPH Grisms}
\subsection{Laboratory measurements}
The diffraction efficiencies of VPH gratings and VPH grisms are measured by using a spectrophotometer with an integral sphere.  The grating is put between suitable prisms to align the incident and diffracted beams for measurements.  If a VPH grating is designed for a VPH grism with ZnSe prisms  of $\alpha=20^{\circ}$, an S-BSL7 (n=1.5 at 1.5\micron) prism for the measurement is $\alpha=51^{\circ}$, for example.

In accordance with the Bragg condition, the efficiency profile is shifted in wavelength due to the tilt of the incident beam into the grisms.  This fact indicates that, since a collimated beam from an astronomical object is tilted as a function of the distance from the field center in the dispersed direction, the wavelengths of the peak efficiency is changed from field to field.  Figure \ref{fig.4} shows relations between the slit position from the field center and the wavelength of the peak efficiency for the VPH grisms of MOIRCS.  It is noted that the calculations of figure \ref{fig.4} made use of measured modulation frequencies (mentioned in subsection 4.3, table 2).  The solid lines indicate calculations in which the incident and diffracted beams in the VPH gratings satisfy Bragg condition.  The beam from the field edge, which is 1'.75 from the center for each focal plane array, is incident on the VPH grism with an angle of $4^{\circ}.6$.  The wavelength of peak efficiencies for the {\it Y-, J-, H-}, and {\it K-}band grisms change by $\sim \pm 0.14$, 0.18, 0.24, and 0.36\micron, respectively.  In other words, the efficiency at a fixed wavelength changes across the field.  For wide-field multislit spectroscopy, observers should be aware of the wavelength shift in the field of view.
 
 \begin{figure}
  \begin{center}
    \FigureFile(131.6mm,50mm){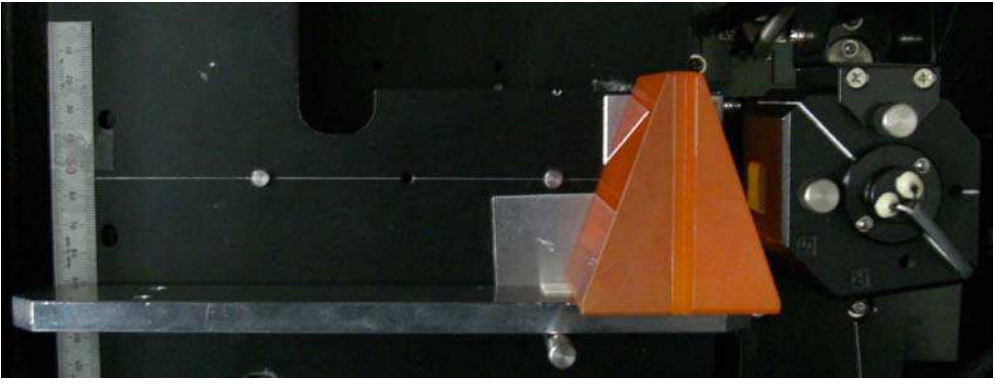}
  \end{center}
  \caption{Measurement of efficiency for a VPH grism by using a spectrophotometer with an integral sphere.}\label{fig.5}
\end{figure}
  
\begin{figure}
  \begin{center}
    \FigureFile(77.3mm,55mm){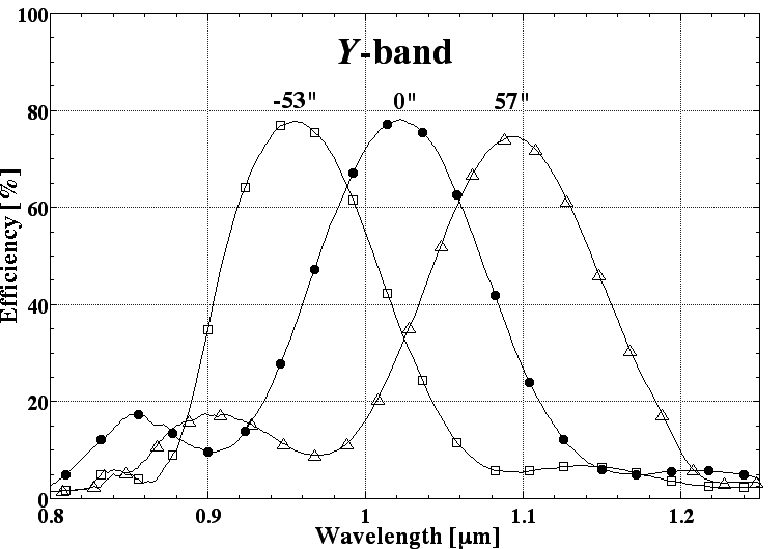}
    \FigureFile(79mm,55mm){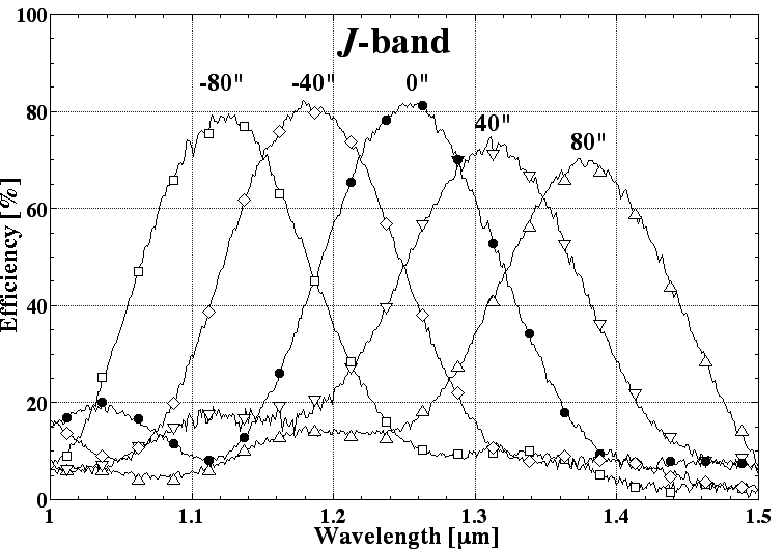}
  \end{center}
  \caption{Dependence of the efficiency for {\it Y-}band (left) and {\it J-}band (right) grisms on the slit position.}\label{fig.6}
\end{figure}

The efficiencies of the VPH grism were also measured with several incident angles by using the spectrophotometer.  The incident angle was determined by a ruler fixed on the spectrophotometer and a bar attached onto base-surface of the grism, as shown in figure \ref{fig.5}.  The efficiencies measured with the nominal incident angles for individual grisms are shown in table\ref{tab.2}.  The peak wavelength of one of the {\it Y-}band grism was found to be shifted by 0.03\micron, and one of the {\it J-}band grism was found to be shifted by 0.014\micron; the {\it Y-} and {\it J-}band grisms are thus installed with tilt of $1^{\circ}.0$ and $0^{\circ}.4$, respectively.  Figure \ref{fig.6} shows the dependence of the efficiency for the {\it Y-} and {\it J-}band grisms on the slit position.  The open circles in figure \ref{fig.4} represent the wavelengths of the peak efficiency of the {\it Y-} and {\it J-}band grisms with several slit positions.  They are in good agreement with the calculated efficiencies (Ichikawa et al. 2008).

\begin{figure}
  \begin{center}
    \FigureFile(77.23mm,55mm){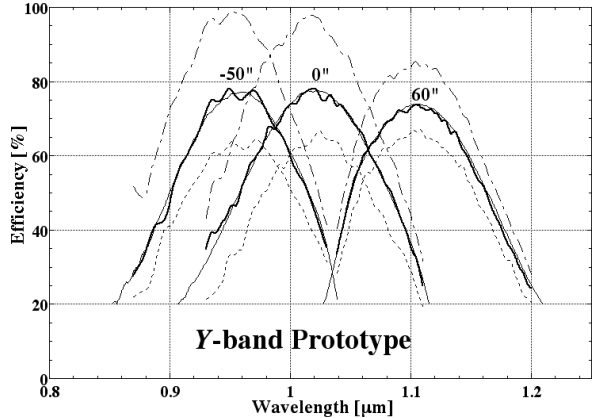}
  \end{center}
  \caption{Polarization properties of a prototype VPH grating for {\it Y-}band grism, $20^{\circ}$ in Bragg-angle, and 1.5 in average refractive index of the VPH grating, measured by using a spectrophotometer with an integration sphere. Bold solid lines indicate efficiencies with $45^\circ$ polarization, thin dot-dashed lines efficiencies with s-polarization, dashed lines efficiencies with p-polarization, and thin solid line efficiencies without polarizer.}\label{fig.7}
\end{figure}

It is noted that the efficiency properties of a VPH grating for s- and p-polarized light --- those electric fields are parallel and orthogonal to the grating lattice, respectively --- are different.  Moreover peaks of the efficiencies for s- and p-polarized light are slightly shifted toward shorter and longer wavelength, respectively, from the peak wavelength of the non-(or $45^{\circ}$- or circular-) polarized light (Baldry et al. 2004).  As shown in figure \ref{fig.7}, the peak efficiency of a prototype VPH grating for {\it Y-}band grism is $\sim75\%$ for the nonpolarized light, while peak efficiencies for s- and p-polarization are 95 and 65\% respectively.
  
The tolerance of the wave-front precision for the MOIRCS VPH grism was given to be $\leq$0.5 waves in rms at each specific wavelength.  That of grisms was measured by a GPI interferometer of Zygo Corp.; we found that wave-front errors of all grisms were small enough for the tolerance.

\subsection{Heat cycle test}
Since the thermal emission from VPH grisms makes serious noise in the infrared for astronomical observations, VPH grisms should be cooled at a cryogenic temperature of $\sim$100 K in MOIRCS.  Before we test the performance of the VPH grating and the VPH grism, we must know if they tolerate under cryogenic circumstances.  Therefore, at first we carefully repeat heat cycles for the VPH grating in a small cryostat at between room temperature and 100 K.  In order not to give large stress during the cycle, we set the cooling and warming speed at 0.1 K per min, which is the slowest speed of the temperature controller that we used (LakeShore/M331), and close to the control speed of MOIRCS.  For unknown reasons, some VPH gratings clouded over after cooling.  Although the transparency was supposed to remain unchanged in the near-infrared, we discarded any cloudy gratings.  We then chose those VPH gratings that did not deteriorate after several heat cycles, and glued ZnSe prisms to the VPH grating.  The thermal cycles were repeated for all of the VPH grisms one by one.

Since the thickness of the resin layer is one of the parameters that define the performance of the grism, as mentioned in section 1, the contraction of resin at low temperature possibly changes the line density and the diffraction efficiency.  The thermal cycle could also deteriorate the performance and the lifetime.  Therefore, it is important to understand the behavior of the VPH gratings at a cryogenic temperature.  Tamura et al. (2006) and Blais-Ouellette et al. (2004) presented the results of cooling tests for VPH gratings, which were made of DCG, at a cryogenic temperature ($\sim100 K$) and found no clear evidence that the diffraction efficiency and blaze wavelength change with a thermal cycle.  To confirm their results, we built a facility for investigating the performance of VPH gratings under cryogenic circumstances, which included a cryogenic chamber, a moving stage, and an IR detector.  The cryostat was fixed on a rotation table, and an IR detector was mounted on a semi-circular table.  The temperature inside the cryostat was monitored by a Pt resister.  The VPH grating was set at the Bragg condition that the incident angle of an input beam to the normal of the grating surface was $\sim30^{\circ}$.  Collimated infrared light exiting from a monochromator and fed into the window illuminates the VPH grating in the cryostat.  The spectral bandwidth was set by the slit width at the exit of the monochromator.  The incident beam was then diffracted by the VPH grating.  The diffracted beam was scanned by the IR detector to measure the efficiency of the spectra as a function of the wavelength.  The measurements were carried out at between room temperature and 120 K (the lowest temperature was limited by the power of the mechanical cooler).  The grating was kept at several temperatures ranging between room and 120 K by a closed-loop controller so as to examine the dependence of the efficiency and blaze wavelengths on the temperature.  We then confirmed the previous results of a small temperature dependence of the efficiency function.  The performance at a cryogenic temperature ($\sim$120 K) was consistent with that at room temperature.  No change in the performance was observed after several thermal cycles within the measurement errors, though it was slightly different among VPH gratings (Ichikawa et al. 2008).

Because it has a small difference for the coefficients of thermal expansion between the substrates of hologram plates (S-BSL7 or S-TIM35) and prisms (ZnSe), the grism would suffer irreparable damage due to a detachment of the glued prisms.  Therefore, to gradually release stress in the VPH grisms during cooling, we repeated thermal cycle four times of cooling down to each targeted temperature of 250 K, 200 K, 150K, and 100 K and then warming up to room temperature at a speed of 0.1 K per min.  After installing them in MOIRCS, we cooled them again to 100 K, and warmed up to room temperature several times.  We found no defect in spite of many thermal cycles, since the grisms were installed in MOIRCS.  This was the first astronomical application of cryogenic VPH grisms.

\subsection{In-situ measurement and test observations}
\begin{figure}
  \begin{center}
    \FigureFile(103.9mm,70mm){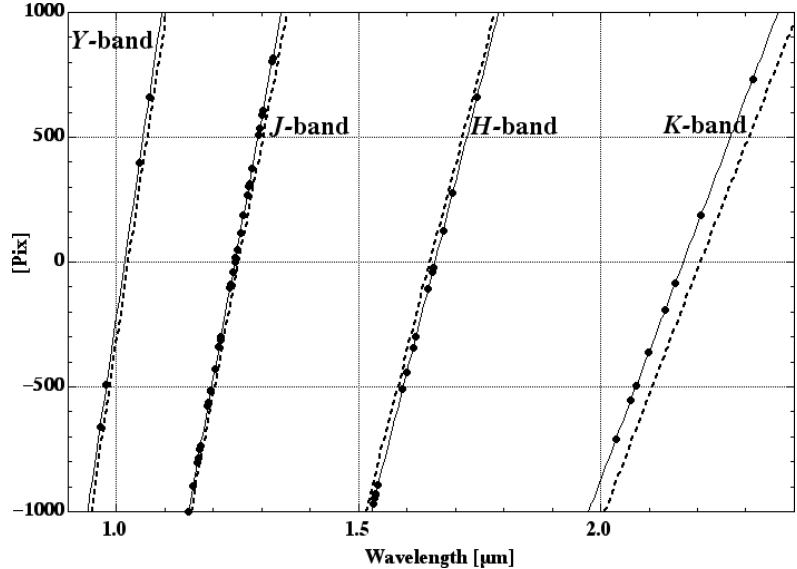}
  \end{center}
 \caption{Spectral positions of VPH grisms on the IR detector of MOIRCS.  Bold dashed lines indicate spectral positions calculated from designed specifications with the central slit.  Filled circles show the measured spectral position of a hollow cathode lamp of Th-Ar with the central slit.  Thin solid lines indicate spectral positions calculated from the measured modulation frequencies with the central slit.}\label{fig.8}
\end{figure}

 To confirm the performance of the VPH grisms for astronomical applications, we mounted the grisms on MOIRCS filter turrets, which are located near to the pupil of MOIRCS.  Figure \ref{fig.8} represents spectral positions of the grisms on the IR detector of MOIRCS obtained by using a hollow cathode lamp of Th-Ar and the dome screen of the Subaru Telescope.  It is noted that channel 2 detector was used for the measurements.  The dispersions for the {\it Y-}, {\it J-}, {\it H-}, and {\it K-}band grisms were found to be 0.077, 0.099, 0.135 and 0.197 nm pixel$^{-1}$, respectively.  The spectral line width ($\sim$2.5 pixels) is well defined by a slit width, 0".3 (2.56 pixels), of MOIRCS.  Therefore we conclude that the resolving powers for the {\it Y-}, {\it J-}, {\it H-}, and {\it K-}band grisms with 0".5 slit were $R=$3180, 3020, 2940, and 2680, respectively (table \ref{tab.2}).  Narrower slit widths give a higher resolving power.  They are in agreement with the original design (table \ref{tab.1}).  Direct vision wavelengths and modulation frequencies of the grisms have been confirmed by the spectra of the emission lamp (figure \ref{fig.8}), as shown in table \ref{tab.2}.  These values must be coincident with those grisms having the same specifications.
   
\begin{table}
  \caption{Measured Performance of MOIRCS VPH grisms (Ichikawa et al. 2008; Nakajima et al. 2008).}\label{tab.2}
  \begin{center}
    \begin{tabular}{lclclclc|c|c|}
      \hline
 Band, & Direct vision & Peak &FWHM & Resolving & Modulation \\
 channel & wavelength & efficiency & (range) & power \footnotemark[$*$] & Frequency \\
 & [\micron] & ($\lambda/\Delta\lambda$) & [\%] & [\micron] & [Periods/mm] \\ \hline
{\it Y}, CH1 & & 78 &0.12 (0.96$\sim$1.08) & & \\
 \hspace{1em} CH2 & 1.017 & 77 & 0.12 (0.96$\sim$1.08)\footnotemark[$\dagger$] & 3180 & 1025 \\
{\it J}, CH1& & 73 & 0.19 (1.16$\sim$1.35)\footnotemark[$\ddagger$] & & \\
\hspace{0.9em} CH2 & 1.245 & 82 & 0.15 (1.18$\sim$1.33) & 3020 & 819 \\
{\it H}, CH1 & & 73 & 0.16 (1.57$\sim$1.73) & & \\
\hspace{1em} CH2 & 1.658 & 70 &  0.18 (1.56$\sim$1.74) & 2940 & 614 \\
{\it K}, CH1 & & 77 & 0.47 (1.97$\sim$2.43) & & \\
 \hspace{1em} CH2 & 2.170 & 80 & 0.46 (1.97$\sim$2.44) & 2680 & 431 \\ \hline
   \multicolumn{4}{@{}l@{}}{\hbox to 0pt{\parbox{85mm}{\footnotesize    
     \par\noindent
 \footnotemark[$*$] Calculated for 0.5" slit.
 \par\noindent
  \footnotemark[$\dagger$] Installed with tilt of $1^{\circ}.0$.
   \par\noindent
     \footnotemark[$\ddagger$] Installed with tilt of $0^{\circ}.4$.
    }\hss}}
    \end{tabular}
  \end{center}
\end{table}

\begin{figure}
  \begin{center}
    \FigureFile(94mm,70mm){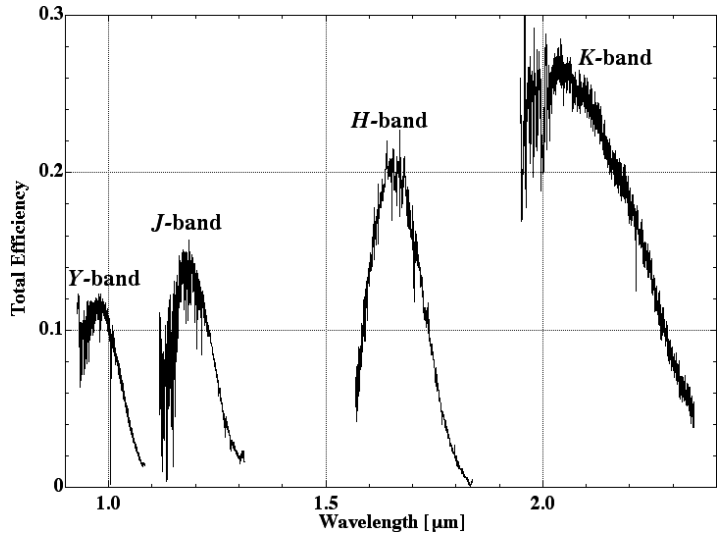}
   \end{center}
  \caption{Total efficiencies of MOIRCS VPH grisms including air mass,  the Subaru Telescope, MOIRCS optics, and the detector.}\label{fig.9}
\end{figure}

We obtained the efficiencies of the grisms by observations of spectroscopic standard stars.  Figure \ref{fig.9} shows the total efficiencies of the MOIRCS VPH grisms including the air mass, the Subaru Telescope, MOIRCS optics, and the detector.  {\it Y-} and {\it K-}band spectra were obtained by slit-less observations, while  {\it J-} and {\it H-}band spectra were obtained by observations with a 0".8 slit.  The channel 2 detector was also used for the observations.  The difference of the total efficiencies among bands is supposed to be mainly caused by the quantum efficiency of the IR detector.  The efficiency of the {\it J-}band is 60\% of {\it K-}band, for example (Suzuki et al. 2008).  Considering the throughput of the MOIRCS optics, we found the efficiency (70$\sim$80\%) to be consistent with laboratory measurements (table \ref{tab.2}).  The filled circles in the figure \ref{fig.4} represent the peak wavelengths of the efficiency obtained by observations of the standard stars.  Shifts from the calculated wavelength indicate the tilts of the VPH grism installations.  The {\it Y-} and {\it J-}band grisms tilt within $0^{\circ}.3$, while the {\it H-} and {\it K-}band grisms tilt with $-1^{\circ}.1$ and $0^{\circ}.9$, respectively, where the plus angle indicates clockwise rotation in the right panel of figure 1.

\section{Astronomical Observations}
\begin{figure}
  \begin{center}
    \FigureFile(90.1mm,80mm){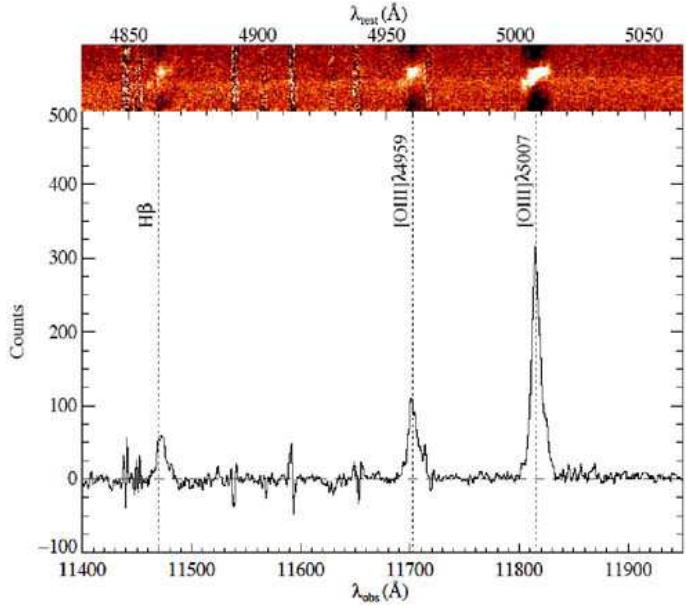}
  \end{center}
  \caption{Star-forming galaxy at z = 1.35930 in GOODS-N taken with {\it J-}band grism on Subaru/MOIRCS.  The emission lines, [O III]  (495.9, 500.7 nm) and H$\beta$ (486.1 nm), are clearly detected and indicated on the 1D spectrum.  The two-dimensional spectrum is also shown in the top panel.  The observed wavelength and rest-frame wavelength are indicated on the bottom and top horizontal axes respectively.}\label{fig.10}
\end{figure}

   The availability of the VPH grisms, which were developed in the present study, is fully demonstrated by an example of real astronomical spectra, as shown in figure \ref{fig.10}, which shows the spectrum of an astronomical object taken with Subaru/MOIRCS equipped with the {\it J-}band grism.  We can see conspicuous emission line features in OH windows.  This is the first astronomical application of cryogenic VPH grisms.  The object was selected from MOIRCS Deep Survey (Kajisawa et al. 2009) as an narrow-band emitter showing excess emission in the NB119 filter (central wevelength of 1.1885\micron\  with 0.0141\micron\  band width), and has been spectroscopically identified by Team Keck Treasury Redshift Survey (TKRS; Wirth et al. 2004).  The redshift derived from the DEIMOS on the Keck telescope was found to be 1.35907 by using [O II] emission line at 372.7 nm.  

The data were taken on 2008 March 30 and 31 with 14 hour integration under fine weather and a 0''.5$\sim$1''.2 seeing condition. A slit width of 0''.8 was used.  Data reduction was carried out with the MCSMDP pipeline (Yoshikawa et al. 2010) including bad-pixel and cosmic-ray rejection, flat-fielding, sky subtraciton, distortion correction, wavelength calibration, and subtraction of residual-sky emission.  Processed frames were then coadded into a 2D spectrum according to appropriate offsets, and a 1D spectrum was extracted with 1".5 aperture.  The typical 1$\sigma$ uncertainty of wavelength calibration is $\simeq$0.05 nm, which corresponds to 12.7 km s$^{-1} $ at the position of [O III]  emission line at 500.7 nm. 

From the MOIRCS spectrum, the redshift of 1.35930 was derived by using [O III] (500.7 nm). 
The difference between TKRS redshift and our redshift translates into a velocity of 30 km s$^{-1}$,  which is within a range of $1\sigma$ uncertainty (38 km s$^{-1}$) of TKRS. From figure \ref {fig.10}, there is an apparent velocity offset in H$\beta$ compared with the expected wavelength derived from the redshift of [O III].  This offset is 6.5 km s$^{-1}$, which is fully within the uncertainty in the wavelength calibration.  The velocity structure is clearly visible in the 2D spectrum, indicating rotating or disturbed motions of ionized gas, which could cause the asymmetric emission line profile seen in the 1D spectrum.  

\begin{figure}
  \begin{center}
    \FigureFile(74.9mm,75mm){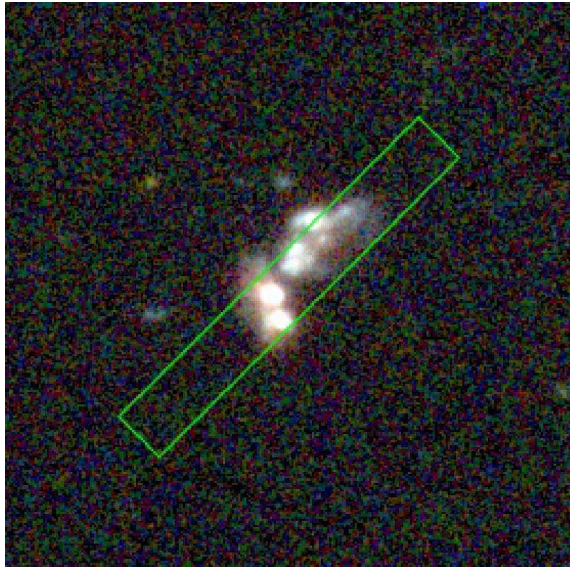}
  \end{center}
  \caption{Composite color image of an object observed with {\it J-}band grism.  The image was generated from {\it HST}/ACS F435W (blue), F606W (green), and F850LP (red) data. The size of the image is 8''$\times$8'' and north is up.  Green rectangle indicates the slit position, while slit length is reduced by a factor 2 for clarity.  Note that the pixel scale of the image is 0''.03 pixel$^{-1}$, while the seeing of the spectra shown in figure \ref{fig.10} was 0''.5$\sim$1''.2.  A companion galaxy is located in the northwest direction from the galaxy.}
      \label{fig.11}
\end{figure}

It must be noted that the galaxy has a companion at $\simeq$ 1'' ($\simeq$ 8.4 kpc)\footnote{$H_0=70 {\rm km\ s}^{-1} {\rm Mpc}^{-1}$, $\Omega_M=0.3$, and $\Omega_\Lambda=0.7$ are used.} in the north-west direction. 
Their appearance in {\it HST}/ACS (figure \ref{fig.11}) images indicates that they are interacting with one another at the same redshift. 
The slit was put along these galaxies by chance, and fainter emission lines, possibly coming from the companion slightly offsetting to a shorter wavelength than those from the main galaxy, are also visible in the 2D spectrum. The corresponding velocity offset is $\simeq$ 270 km s$^{-1} $. The physical separation and velocity difference between them confirm that they are a real interacting system. 
 
By combining the {\it J-}band VPH spectrum of MOIRCS with DEIMOS spectrum, it could be possible to derive the gas-phase oxygen abundance from strong indicators, such as $R_{23}$ and $O_{32}$ (e.g., Pagel et al. 1979; Kobulnicky \& Kewley 2004).  A detailed analysis of the system will be presented elsewhere.

\section{Conclusions}
The high-dispersion VPH grisms with ZnSe prisms achieve very high efficiencies. The reasons are of follows: a VPH grating inheres the high performance; the ZnSe prism gives a large optical pass difference compared with a prism of conventional optical glass; and the hologram resin have less absorption in the near infrared wavelength up to 2.7\micron.  Moreover, the grisms are resistant to a cryogenic temperature.  However, the user should notice that a VPH grism has some dependence of the wavelength of the peak efficiency on the slit position.  From another viewpoint, the peak wavelength as the user's request can be adjusted by the slit position.

Common use for the {\it J-} and {\it H-}band grisms started in the second semester of 2008.   The performance has been found to be very consistent with the results, as evaluated in the laboratory, since the grisms have been in commission starting from February 2008.  Although the grisms were originally developed for high-resolution spectroscopy of galaxies or distant gamma-ray bursts, these can also be used for general targets to obtain high-resolution spectra between 0.9 and 2.5\micron.  Common use for the {\it Y-}band grisms started in the second semester of 2010, and that for the {\it K-}band grisms will start in the first semester of 2011.

\bigskip
We greatly appreciate M. Kawabata and T. Teranishi of Nippon Paint Co. Ltd. for supplying the hologram resin.  We thank T. Harashima, Y. Okura of Soma Optics Ltd., Y. Taniguchi and T. Murata of Tohoku University for their contributions on VPH grating fabrications and measurements.  Y. Komai, E. Watanabe, and member of Kodate laboratory of Japan Women's University (JWU) gave us valuable opinions on VPH grating fabrications.  T. Koyano of Tohoku University kindly made parts of the cryogenic Dewar Vessel.  Continuous advice and encourage by MOIRCS builders are also acknowledged.  T. Nagao of Ehime University helped us to select targets for commissioning observations.  We thank M. Ouchi of The University of Tokyo for providing the {\it J-}band VPH spectra before publication.  NE appreciate A. Makinouchi of RIKEN, F. Kajino of Konan University, and S. Sato and M. Hori of Nagoya University for their supports of our research.  We utilize facility of RIKEN, JWU and the Advanced Technology Center of National Astronomical Observatory of Japan (NAOJ) for manufacture of the VPH grisms and for grating measurements.  This research was supported by a grant-in-aid of NAOJ for R\&D of the Subaru Telescope instruments, a grant-in-aid of RIKEN for practical use of research results, and partly supported from the Ministry of Education, Culture, Sports, Science and Technology by Grant-in-Aid for Scientific Research (B)(1998-2001, 09559018; 2002-2004,14300059) and for Specially Promoted Research (2007-2010, 19035003; 2007-2010, 19047003).

\end{document}